\newcommand\SEKIusersusepackages
{%\usepackage{pipifax}
}

\newcommand\SEKIREVIEWSTATUS{1}
\newcommand\SEKISUBEDITOR
{Erica Melis 
\\FR Informatik, Universit\"at des Saarlandes, D--66123 Saarbr\"ucken, Germany
\\E-mail: {\tt melis@activemath.org}
\\WWW: \url{http://www.activemath.org/~melis/}
}

\newcommand\SEKIREVIEWER
{\\Claus-Peter Wirth
\\E-mail: \url{wirth@logic.at}
\\WWW: \url{http://www.ags.uni-sb.de/~cp}
\\}

\newcommand\SEKIREFEREEPROCESS
{
}

\newcommand\SEKIDOCUMENTCLASS{0}
\newcommand\SEKIYEAR{2008}

\newcommand\SEKINUMBEROFYEAR{01}
\newcommand\SEKITYPE{0}
\newcommand\SEKIPUBLISHEDTYPE{0}
\newcommand\SEKIPUBLISHEDAS
{}
\date{October 28, 2008}

\title{Automating Access Control Logics in Simple Type Theory with LEO-II}

\author
{
Christoph Benzm\"uller \\ 
FR Informatik, Saarland Univiversity, Germany\\
{\tt chris@ags.uni-sb.de}\\
}

\input seki-deckblatt-2

\usepackage{mathptmx}       % selects Times Roman as basic font
\usepackage{helvet}         % selects Helvetica as sans-serif font
\usepackage{courier}        % selects Courier as typewriter font
\usepackage{type1cm}        % activate if the above 3 fonts are
\usepackage{makeidx}         % allows index generation
\usepackage{graphicx}        % standard LaTeX graphics tool
\usepackage{multicol}        % used for the two-column index
\usepackage[bottom]{footmisc}% places footnotes at page bottom

\usepackage{latexsym}
\usepackage{longtable}
\usepackage{wrapfig}
\usepackage{verbatim}
\usepackage{url}
\usepackage{xspace}

\newenvironment{proof}{{\bf Proof: }}{\qed}

\def\lambdot{\rule{0.6mm}{0.6mm}\hspace{0.4ex}} 
\def\all#1{\forall #1\lambdot}
\def\exi#1{\exists #1\lambdot}
\def\lam#1{\lambda #1\lambdot}

\def\modal#1{\mbox{\blue \boldmath $#1$}}
\def\mfalse{\modal\bot}
\def\mtrue{\modal\top}
\def\mnot{\modal\neg\,}
\def\mor{\,\modal\vee\,}
\def\mand{\,\modal\wedge\,}
\def\mimpl{\,\modal\supset\,}

\def\mball#1{\modal\Box_{#1}\,}

\def\iclfalse{\bot}
\def\icltrue{\top}
\def\iclor{\,\vee\,}
\def\icland{\,\wedge\,}
\def\iclimpl{\,\supset\,}
\def\iclsays{\,\,{\texttt{says}}\,\,}
\def\iclimplprinc{\,\Longrightarrow\,}

\newenvironment{ednote}{\begin{svgraybox}\noindent\textbf{EdNote:}\par}
  {\end{svgraybox}}

\newcommand{\database}[1]{
\rput(0,1.5){\rnode{h1}{\psellipse(1,0)(1,0.3)}}  %% first pair places the centre, second determines shape
\rput(0,0){\psline{-}(0,1.5)}
\rput(2,0){\psline{-}(0,1.5)}
\rput(0,0){\rnode{h2}{\psline[linearc=0.7]{-}(0,0)(0.2,-0.12)(0.6,-0.23)(1,-0.26)(1.4,-0.23)(1.8,-0.12)(2,0)}}
}

\def\modal#1{\mbox{\boldmath $#1$}}
\def\mfalse{\modal\bot}
\def\mtrue{\modal\top}
\def\mnot{\modal\neg\,}
\def\mor{\,\modal\vee\,}
\def\mand{\,\modal\wedge\,}
\def\mimpl{\,\modal\supset\,}

\def\mball#1{\modal\Box_{#1}\,}

\def\Metaeq{=}

\def\ambnormform#1{{#1}\hspace*{-1.1ex}\downarrow_{\kern-.2em\scriptscriptstyle *}} % ``ambiguous'' normal form
\def\eqb{{\Metaeq_{\beta}}}

\def\eqbe{{\Metaeq_{\beta\eta}}}

\def\stt{\textit{STT}\xspace}
\newcommand\ttremoveclash{STT}
\def\lm{\textit{ML}\xspace}

\def\deq{{\;\colon\kern-.1em=\;}}
\makeatletter\renewcommand{\ednote}[2][]{\ed@note{#2}{E}{#1}}\makeatother

\def\qed{\hfil\null\nobreak\hfill{q.e.d}\par\smallskip}

\begin{document}
\makecover

%------------------------------------------------------------------------- 
\maketitle

\begin{abstract}%
  Garg and Abadi recently proved that prominent access control logics
  can be translated in a sound and complete way into modal logic
  S4. We have previously outlined how normal multimodal logics,
  including monomodal logics K and S4, can be embedded in simple type
  theory (which is also known as higher-order logic) and we have
  demonstrated that the higher-order theorem prover LEO-II can
  automate reasoning in and about them. In this paper we combine these
  results and describe a sound and complete embedding of different
  access control logics in simple type theory. Employing this
  framework we show that the off the shelf theorem prover LEO-II can
  be applied to automate reasoning in prominent access control logics.
\end{abstract}

\section{Introduction}
The provision of effective and reliable control mechanisms for
accessing resources is an important issue in many areas. In computer
systems, for example, it is important to effectively control the
access to personalized or security critical files.

A prominent and successful approach to implement access control relies
on logic based ideas and tools. Abadi's article \cite{Abadi03}
provides a brief overview on the frameworks and systems that have been
developed under this approach.  Garg and Abadi recently showed that
several prominent access control logics can be translated into modal
logic S4 \cite{GargAbadi08}. They proved that this translation is
sound and complete.

We have previously shown \cite{B9} how multimodal logics can be elegantly
embedded in simple type theory (\stt) \cite{Church40,Andrews2002a} --- which is widely also known as higher-order logic (HOL). 
We have also demonstrated that proof problems in and about multimodal
logics can be effectively automated with the higher theorem prover
LEO-II.

In this paper we combine the above results and show that different
access control logics can be embedded in \stt, which has a well
understood syntax and semantics
\cite{Henkin50,Andrews72a,Andrews72b,BBK04}.

The expressiveness of \stt furthermore enables the encoding of the
entire translation from access control logic input syntax to \stt in
\stt itself, thus making it as transparent as possible.  Our embedding
furthermore demonstrates that prominent access control logics as well
as prominent multimodal logics can be considered and treated as
natural fragments of \stt.

Using our embedding, reasoning in and about access control logic
can be automated in the higher-order theorem prover LEO-II \cite{C26}.
Since LEO-II generates proof objects the entire translation and
reasoning process is in principle accessible for independent proof 
checking.

% For this, only the
% translation encodings are modified and extended while the simple
% type theory prover remains unchanged.
% Our approach is absolutely transparent and entirely verifiable: Access
% control logics problems are originally encoded using the regular
% access control logics syntax.  The translation into simple type theory
% is then realized via definition expansion in LEO-II and these
% expansion steps are explicitly represented in LEO-II's proof objects.
% LEO-II then searches for a proof (actually a refutation proof) for the
% fully expanded problems and, if successful, returns a proof object.
% This proof object could be exploited for independend proof
% verification and for generating verbal proof explanations.

This paper is structured as follows: Section \ref{prelim} reviews
background knowledge and Section \ref{gargabadi} outlines the
translation of access control logics into modal logic S4 as proposed
by Garg and Abadi \cite{GargAbadi08}. Section \ref{benzpaulson}
restricts the general embedding of multimodal logics into \stt \cite{B9}
to an embedding of monomodal logics K and S4 into \stt and proves its
soundness and completeness.  These results are combined in Section \ref{combined} in
order to obtain a sound and complete embedding of access control logics into
\stt. Moreover, we present some first empirical evaluation of the
approach with the higher-order automated theorem prover
LEO-II. Section \ref{conc} concludes the paper.

\section{Preliminaries} \label{prelim}
We assume familiarity with the syntax and semantics and of multimodal logics 
and simple type theory  and only briefly review the most important notions.

%\subsection{Multimodal Logics}
The multimodal logic language $\lm$ is defined by
$$ s,t ::= p | \mnot s | s \mor t | \mball{r} s $$
where $p$ denotes atomic primitives and $r$ denotes accessibility
relations (distinct from $p$).  Other logical connectives can be
defined from the chosen ones in the usual way.

A Kripke frame for $\lm$ is a pair $\langle W,(R_r)_{r\in
  I}\rangle$, where $W$ is a non-empty set (called possible worlds),
and the $R_r$ are binary relations on $W$ (called accessibility
relations). A Kripke model for $\lm$ is a triple $\langle
W,(R_r)_{r\in I},\models\rangle$, where $\langle W,(R_r)_{r\in
  I}\rangle$ is a Kripke frame, and $\models$ is a satisfaction
relation between nodes of $W$ and formulas of $\lm$ satisfying:
$w\models \mnot s$ if and only if $w\not\models s$, $w\models s \mor
t$ if and only if $w\models s$ or $w\models t$, $w \models \mball{r}
s$ if and only if for all $u$ with $R_r(w,u)$ holds $u \models s$. The
satisfaction relation $\models$ is uniquely determined by its value on
the atomic primitives $p$.  A formula $s$ is valid in a Kripke model
$\langle W,(R_r)_{r\in I},\models\rangle$, if $w\models s$ for all $w
\in W$. $s$ is valid in a Kripke frame $\langle W,(R_r)_{r\in
  I}\rangle$ if it is valid in $\langle W,(R_r)_{r\in
  I},\models\rangle$ for all possible $\models$.  If $s$ is valid for
all possible Kripke frames $\langle W,(R_r)_{r\in I}\rangle$ then $s$
is called valid and we write $\models^{K} s$.  $s$ is called $S4$-valid (we write
$\models^{S4} s$) if it is
valid in all reflexive, transitive Kripke frames $\langle
W,(R_r)_{r\in I}\rangle$, that is, Kripke frames with only reflexive
and transitive relations $R_r$.

% \subsection{Simple Type Theory}
Classical higher-order logic or simple type theory \stt
\cite{Andrews2002a,Church40} is a formalism built on top of the simply
typed $\lambda$-calculus. The set~${\cal T}$ of simple types is
usually freely generated from a set of basic types $\{o, \iota\}$
(where $o$ denotes the type of Booleans) using the function type
constructor $\rightarrow$.

The simple type theory language \stt is defined by
($\alpha,\beta,o\in{\cal T}$):
\begin{eqnarray*} 
& s,t ::= \\
& p_\alpha | X_\alpha | (\lam{X_\alpha} s_\beta)_{\alpha\rightarrow\beta} | (s_{\alpha\rightarrow\beta}\, t_\alpha)_\beta | (\neg_{o\rightarrow o}\, s_o)_o | (s_o \vee_{o\rightarrow o \rightarrow o} t_o)_o | (\Pi_{(\alpha\rightarrow o)\rightarrow o}\, s_{\alpha\rightarrow o})_o 
\end{eqnarray*}
$p_\alpha$ denotes typed constants and $X_\alpha$ typed variables
(distinct from $p_\alpha$) .  Complex typed terms are constructed via
abstraction and application.  Our logical connectives of choice are
$\neg_{o\rightarrow o}$, $\lor_{o\rightarrow o\rightarrow o}$ and
$\Pi_{(\alpha\rightarrow o)\rightarrow o}$ (for each type $\alpha$).
From these connectives, other logical connectives can be defined in
the usual way. We often use binder notation $\all{X_\alpha} s$ for
$(\Pi_{(\alpha\rightarrow o)\rightarrow o}(\lam{X_\alpha} s_o))$.  We
denote substitution of a term $A_\alpha$ for a variable $X_\alpha$ in
a term $B_\beta$ by $[A/X]B$.  Since we consider $\alpha$-conversion
implicitly, we assume the bound variables of $B$ avoid variable
capture.  Two common relations on terms are given by $\beta$-reduction
and $\eta$-reduction.  A $\beta$-redex $(\lam{X}s)t$ $\beta$-reduces
to $[t/X]s$.  An $\eta$-redex $(\lam{X}s X)$ where variable $X$ is not
free in $s$, $\eta$-reduces to $s$.  We write $s\eqb t$ to mean $s$
can be converted to $t$ by a series of $\beta$-reductions and
expansions.  Similarly, $s\eqbe t$ means $s$ can be converted to $t$
using both $\beta$ and $\eta$.
% For each $s\in L$ there is a 
% unique $\beta$-normal form (denoted $\Bnormform{s}$)
% and a unique $\beta\eta$-normal form (denoted $\Benormform{s}$).
% From this fact we know $s\eqb t$ ($s\eqbe t$) iff
% $\Bnormform s\Metaeq\Bnormform t$
% ($\Benormform s\Metaeq\Benormform t$).
 
Semantics of \stt is well understood and thoroughly
documented in the literature
\cite{BBK04,Andrews72b,Andrews72a,Henkin50}; our summary below is adapted
from Andrews \cite{sep-type-theory-church}.

 A frame is a collection $\{D_\alpha\}_{\alpha\in{\cal T}}$ of
 nonempty domains (sets) $D_\alpha$, such that $D_o = \{T, F\}$ (where
 $T$ represents truth and $F$ represents falsehood).  The
 $D_{\alpha\rightarrow\beta}$ are collections of functions mapping
 $D_\alpha$ into $D_\beta$. The members of $D_\iota$ are called
 individuals. An interpretation is a tuple $\langle
 \{D_\alpha\}_{\alpha\in{\cal T}}, I \rangle$ where function $I$ maps
 each typed constant $c_\alpha$ to an appropriate element of
 $D_\alpha$, which is called the denotation of $c_\alpha$ (the denotations of
$\neg$,
 $\vee$ and $\Pi$ are always chosen as intended). A variable
 assignment $\phi$ maps variables $X_\alpha$ to elements in
 $D_\alpha$.  An interpretation $\langle \{D_\alpha\}_{\alpha\in{\cal
     T}}, I \rangle$ is a Henkin model (general model) if and only if
 there is a binary function ${\cal V}$ such that ${\cal V}_\phi\,
 s_\alpha \in D_\alpha$ for each variable assignment $\phi$ and term
 $s_\alpha\in L$, and the following conditions are satisfied for all
 $\phi$ and all $s,t\in L$: (a) ${\cal V}_\phi X_\alpha = \phi
 X_\alpha$, (b) ${\cal V}_\phi\, p_\alpha = I p_\alpha$, (c) ${\cal
   V}_\phi (s_{\alpha\rightarrow\beta}\, t_\alpha) = ({\cal V}_\phi\,
 s_{\alpha\rightarrow\beta}) ({\cal V}_\phi t_\alpha$), and (d) ${\cal
   V}_\phi (\lam{X_\alpha} s_\beta)$ is that function from $D_\alpha$
 into $D_\beta$ whose value for each argument $z\in D_\alpha$ is
 ${\cal V}_{[z/X_\alpha],\phi} s_\beta$, where ${[z/X_\alpha],\phi}$ is that variable assignment
 such that $({[z/X_\alpha],\phi})X_\alpha = z$ and $({[z/X_\alpha],\phi}) Y_\beta = \phi Y_\beta$ if
 $Y_\beta \not= X_\alpha$.\footnote{Since $I\/ \neg$, $I\/
   \vee$, and $I\/ \Pi$ are always chosen as intended, we have ${\cal
     V}_\phi\, (\neg s) = T$ iff ${\cal V}_\phi\, s = F$ , ${\cal
     V}_\phi\, (s \vee t) = T$ iff ${\cal V}_\phi\, s = T$ or ${\cal
     V}_\phi\, t = T$, and ${\cal V}_\phi\, (\all{X_\alpha} s_o) = {\cal
     V}_\phi\, (\Pi^\alpha(\lam{X_\alpha} s_o)) = T$ iff for all $z\in
   D_\alpha$ we have ${\cal
     V}_{[z/X_\alpha],\phi}\, s_o = T$. Moreover, we have ${\cal V}_\phi\, s = {\cal V}_\phi\, t$ whenever $s\eqbe t$.}

 If an interpretation $\langle \{D_\alpha\}_{\alpha\in{\cal T}}, I
 \rangle$ is a Henkin model, the function ${\cal V}_\phi$ is uniquely
 determined. An interpretation $\langle \{D_\alpha\}_{\alpha\in{\cal
     T}}, I \rangle$ is a standard model if and only if for all
 $\alpha$ and $\beta$, $D_{\alpha\rightarrow\beta}$ is the set of all
 functions from $D_\alpha$ into $D_\beta$. Each standard model is
 also a Henkin model.

 We say that formula $A\in L$ is valid in a model $\langle
 \{D_\alpha\}_{\alpha\in{\cal T}}, I \rangle$ if an only if ${\cal
   V}_\phi A = T$ for every variable assignment $\phi$. A model for a
 set of formulas $H$ is a model in which each formula of $H$ is valid.

 A formula $A$ is Henkin-valid (standard-valid) if and only if $A$ is
 valid in every Henkin (standard) model.  Clearly each formula which
 is Henkin-valid is also standard-valid, but the converse of this
 statement is false. We write $\models^{\ttremoveclash} A$ if $A$ is Henkin-valid
 and we write $\Gamma \models^{\ttremoveclash} A$ if $A$ is valid in all Henkin
 models in which all formulas of $\Gamma$ are valid.

\vfill\pagebreak

\section{Translating Access Control Logic to Modal Logic}
\label{gargabadi}
%\subsection{Access Control Logic $ICL$} 
The access control logic $ICL$ studied by Garg and Abadi \cite{GargAbadi08}
is defined by
$$ s ::= p\,|\,s_1 \icland s_2\,|\,s_1 \iclor s_2\,|\,s_1 \iclimpl
s_2\,|\,\iclfalse\,|\,\icltrue\,|\,A \iclsays s $$ $p$ denotes atomic
propositions, $\icland$, $\iclor$, $\iclimpl$, $\iclfalse$ and
$\icltrue$ denote the standard logical connectives, and $A$ denotes
principals, which are atomic and distinct from the atomic propositions
$p$. Expressions of the form \texttt{A says s}, intuitively mean that
$A$ \texttt{asserts} (or \texttt{supports}) $s$.  $ICL$ inherits all
inference rules of intuitionistic propositional logic. The logical
connective \texttt{says} satisfies the following axioms:

\begin{center}
\begin{tabular}{l@{\hspace*{1em}}l}
 $\vdash s \iclimpl (A \iclsays s)$ & (unit) \\
 $\vdash (A \iclsays (s \iclimpl t)) \iclimpl (A \iclsays s) \iclimpl (A \iclsays t)$ & (cuc) \\
 $\vdash (A \iclsays A \iclsays s) \iclimpl (A \iclsays s)$ & (idem) \\
\end{tabular}
\end{center}

\begin{example}[from \cite{GargAbadi08}]
  We consider a file-access scenario with an administrating principal
  \texttt{admin}, a user \texttt{Bob}, one file \texttt{file1}, and
  the following policy:
\begin{enumerate}
\item If \texttt{admin} says that \texttt{file1} should be deleted,
  then this must be the case.
\item \texttt{admin} trusts Bob to decide whether \texttt{file1}
  should be deleted.
\item \texttt{Bob} wants to delete \texttt{file1}.
\end{enumerate}
This policy can be encoded  in $ICL$ as follows:
\begin{center}
\begin{tabular}{l@{\hspace*{1em}}l}
  $(\texttt{admin} \iclsays \texttt{deletefile1}) \iclimpl \texttt{deletefile1}$ & (1.1)\\
  $\texttt{admin} \iclsays ((\texttt{Bob} \iclsays \texttt{deletefile1}) \iclimpl \texttt{deletefile1})$ & (1.2)\\
  $\texttt{Bob} \iclsays \texttt{deletefile1}$ &  (1.3)
\end{tabular}
\end{center}

The question whether \texttt{file1} should be deleted in this
situation corresponds to proving  $\texttt{deletefile1}$ (1.4),
which follows from (1.1)-(1.3), (unit), and (cuc).
\end{example}

Garg and Abadi \cite{GargAbadi08} propose the following mapping
$\lceil . \rceil$ of $ICL$ formulas into modal logic S4 formulas
(similar to G\"odels translation from intuitionistic logic to S4
\cite{Goedel33}, but providing a mapping for the additional connective
$\iclsays$; we refer to Garg and Abadi \cite{GargAbadi08} for a brief discussion of the intuition of the mapping of ${\texttt{says}}$).
 
\begin{minipage}{\textwidth}
\begin{minipage}{.45\textwidth}
\begin{eqnarray*}
  \lceil p \rceil & = & \Box p \\
  \lceil s \wedge t \rceil & = & \lceil s \rceil \wedge \lceil t \rceil \\
  \lceil s \vee t \rceil & = & \lceil s \rceil \vee \lceil t \rceil \\
  \lceil s \supset t \rceil & = & \mball{} (\lceil s \rceil \supset \lceil t \rceil) \\
\end{eqnarray*}
\end{minipage}
\begin{minipage}{.45\textwidth}
\begin{eqnarray*}
  \lceil \mtrue \rceil & = & \top \\
  \lceil \mfalse \rceil & = & \bot \\
  \lceil A \, \, {\texttt{says}} \, \, s \rceil & = & \Box (A \vee \lceil s \rceil ) 
\end{eqnarray*}
\end{minipage}
\end{minipage}

%\subsection{Access Control Logic ${ICL}^{\iclimplprinc}$}
Logic $ICL^{\iclimplprinc}$ extends $ICL$ by a \textit{speaks-for\/}
operator (represented by $\iclimplprinc$) which satisfies the
following axioms:
\begin{center}
\begin{tabular}{l@{\hspace*{1em}}l}
  $\vdash  A \iclimplprinc A$ & (refl) \\
  $\vdash  (A \iclimplprinc B) \iclimpl  (B \iclimplprinc C) \iclimpl (A \iclimplprinc C)$ & (trans) \\
  $\vdash  (A \iclimplprinc B) \iclimpl  (A \iclsays s) \iclimpl (B \iclsays s)$ & (speaking-for) \\
  $\vdash (B \iclsays (A \iclimplprinc B)) \iclimpl (A \iclimplprinc B)$ & (handoff)  
\end{tabular}
\end{center}
The use of the new $\iclimplprinc$ operator is illustrated by the following modification of Example 1.
\begin{example}[from \cite{GargAbadi08}]
  \texttt{Bob} delegates his authority to delete \texttt{file1} to
  \texttt{Alice} (see (2.3)), who now wants to delete \texttt{file1}.
\begin{center}
\begin{tabular}{l@{\hspace*{1em}}l}
$(\texttt{admin} \iclsays \texttt{deletefile1}) \iclimpl \texttt{deletefile1}$ & (2.1)\\
$\texttt{admin} \iclsays ((\texttt{Bob} \iclsays \texttt{deletefile1}) \iclimpl \texttt{deletefile1})$ & (2.2)\\
$\texttt{Bob} \iclsays \texttt{Alice} \iclimplprinc \texttt{Bob}$ & (2.3) \\
$\texttt{Alice} \iclsays \texttt{deletefile1}$ &  (2.4)
\end{tabular}
\end{center}
Using these facts and (handoff) and (speaking-for) one can prove
$\texttt{deletefile1}$ (2.5)
\end{example}
The translation of $ICL^{\iclimplprinc}$ into S4 extends the translation from
$ICL$ to S4 by
\begin{eqnarray*}
\lceil A \iclimplprinc B \rceil & = & \Box (A \iclimpl B) 
\end{eqnarray*}

%\subsection{Access Control Logic $\textit{ICL}^{B}$}
Logic $ICL^{B}$ differs from $ICL$ by allowing that principals may contain
Boolean connectives ($a$ denotes atomic principals distinct from atomic propositions):
$$ A,B ::= a\,|\,A \icland B\,|\,A \iclor B\,|\,A \iclimpl
B\,|\,\iclfalse\,|\,\icltrue\,$$ 
$ICL^{B}$ satisfies the following
additional axioms:
\begin{center}
\begin{tabular}{l@{\hspace*{1em}}l}
  $\vdash  (\iclfalse \iclsays s) \iclimpl s$ & (trust) \\
  If $A\equiv \icltrue$ then $\vdash A \iclsays \iclfalse$ & (untrust) \\
  $\vdash  ((A \iclimpl B) \iclsays s) \iclimpl  (A \iclsays s) \iclimpl (B \iclsays s)$ & (cuc') \\
\end{tabular}
\end{center}
Abadi and Garg show that the \texttt{speaks-for} operator from
$ICL^{\iclimplprinc}$ is definable in $ICL^{B}$. The use of $ICL^{B}$ is illustrated by the following modification of Example 1.
\begin{example}[from \cite{GargAbadi08}] \texttt{admin} is trusted on \texttt{deletefile1}
and its consequences (3.1). (3.2) says that \texttt{admin} further delegates
this authority to \texttt{Bob}.  
\begin{center}
\begin{tabular}{l@{\hspace*{1em}}l}
  $(\texttt{admin} \iclsays \iclfalse) \iclimpl \texttt{deletefile1}$ & (3.1)\\
  $\texttt{admin} \iclsays ((\texttt{Bob} \iclimpl \texttt{admin}) \iclsays \texttt{deletefile1})$ & (3.2)\\
  $\texttt{Bob} \iclsays \texttt{deletefile1}$ &  (3.3)
\end{tabular}
\end{center}
Using these facts and the available
axioms one can again prove $\texttt{deletefile1}$ (3.4).
\end{example}
The translation of $ICL^{B}$ into S4 is the same as the translation from $ICL$
to S4. However, the mapping $\lceil A \, \, {\texttt{says}} \, \, s \rceil =  \Box (A \vee \lceil s \rceil)$
now guarantees that Boolean principal expressions $A$ are mapped
one-to-one to Boolean expressions in S4.

Garg and Abadi prove their translations sound and complete:
% The above translations of access control logics $ICL$, $ICL^{\Rightarrow}$ and $ICL^{B}$ into 
% modal logic S4 are sound and complete \cite{GargAbadi08}.

\begin{theorem}[Soundness and Completeness]\label{thm:gargabadi}
$\vdash s$ in $ICL$ (resp. $ICL^{\Rightarrow}$ and $ICL^{B}$)  if and only if $\vdash \lceil s \rceil$ in S4. 
\end{theorem}
\begin{proof} See Theorem 1 (resp. Theorem 2 and Theorem 3) of Garg and Abadi \cite{GargAbadi08}.
\end{proof}

\section{Embedding Modal Logic in Simple Type Theory} \label{benzpaulson}
Embeddings of modal logics into higher-order logic have not yet been
widely studied, although multimodal logic can be regarded as a natural
fragment of \stt.  Gallin
\cite{Gallin75} appears to mention the idea first. He presents an
embedding of modal logic into a 2-sorted type theory. This idea is
picked up by Gamut \cite{gamut:91.2} and a related embedding has
recently been studied by Hardt and Smolka
\cite{hardt07:_higher_order_syntax_and_satur}.
Carpenter~\cite{carpenter98:_type} proposes to use lifted connectives,
an idea that is also underlying the embeddings presented by
Merz~\cite{merz99:_yet_anoth_encod_of_tla_in_isabel},
Brown~\cite{brown05:_encod_hybrid_logic_in_higher_order_logic},
Harrison \cite[Chap.~20]{harrison06:_hol_light_tutor_for_version}, and
Kaminski and Smolka~\cite{DBLP:conf/cade/KaminskiS08}.

In \cite{B9} we pick up and extend the embedding of multimodal logics
into \stt as studied by
Brown~\cite{brown05:_encod_hybrid_logic_in_higher_order_logic}.  The
starting point is a characterization of multimodal logic formulas as
particular $\lambda$-terms in \stt.  A distinctive
characteristic of the encoding is that the definiens of the
$\mball{R}$ operator $\lambda$-abstracts over the accessibility
relation $R$. As is shown in \cite{B9} this supports the formulation
of meta properties of encoded multimodal logics such as the
correspondence between certain axioms and properties of the
accessibility relation $R$. And some of these meta properties can be
efficiently automated within our higher-order theorem prover LEO-II.

The general idea of this encoding is very simple: Choose base type
$\iota$ and let this type denote the set of all possible
worlds. Certain formulas of type $\iota\rightarrow o$ then correspond
to multimodal logic expressions, whereas the modal operators $\mnot$,
$\mor$, and $\mball{r}$ itself become $\lambda$-terms of type
${(\iota\rightarrow o)\rightarrow(\iota\rightarrow o)}$,
${(\iota\rightarrow o)\rightarrow(\iota\rightarrow
  o)\rightarrow(\iota\rightarrow o)}$, and \
$(\iota\rightarrow\iota\rightarrow o)\rightarrow(\iota\rightarrow
o)\rightarrow(\iota\rightarrow o)$ respectively. Intuitively, a
multimodal formula of type $\iota\rightarrow o$ denotes the set of
worlds in which it is true.

The mapping $\lfloor.\rfloor$ translates formulas of multimodal logic
$\lm$ into terms of type ${\iota\rightarrow o}$ in \stt:
\begin{center}
\begin{minipage}{\textwidth}
  \begin{minipage}{.45\textwidth}
    \begin{eqnarray*}
      \lfloor p \rfloor & = & p_{\iota\rightarrow o} \\
      \lfloor r \rfloor & = & r_{\iota\rightarrow\iota\rightarrow o} \\
      \lfloor \mnot\,s \rfloor & = & \lam{X_\iota}\neg (\lfloor s \rfloor \,X) \\
      \lfloor s\,\mor\,t \rfloor & = & \lam{X_\iota} (\lfloor s \rfloor \,X) \vee (\lfloor t \rfloor \,X) \\
      \lfloor \mball{r}\,s \rfloor & = & \lam{X_\iota} \all{Y_{\iota}} (\lfloor r \rfloor\,X\,Y) \Rightarrow (\lfloor s \rfloor\,Y) \\
    \end{eqnarray*} 
  \end{minipage} \hfill 
  \begin{minipage}{.6\textwidth}
    \begin{eqnarray*}
      | p | & = & p_{\iota\rightarrow o} \\
      | r | & = & r_{\iota\rightarrow\iota\rightarrow o} \\
      | \mnot | & = & \lam{A_{\iota\rightarrow o}}\lam{X_\iota}\neg (A\,X) \\
      | \mor | & = & \lam{A_{\iota\rightarrow o}} \lam{B_{\iota\rightarrow o}} \lam{X_\iota} (A\,X) \vee (B\,X) \\
      | \mball{} | & = & \lam{R_{\iota\rightarrow\iota\rightarrow o}} \lam{A_{\iota\rightarrow o}} \\
      & & \hspace*{1cm} \lam{X_\iota} \all{Y_{\iota}} (R\,X\,Y) \Rightarrow (A\,Y) 
    \end{eqnarray*}
  \end{minipage}
\end{minipage}
\end{center}
The expressiveness of \stt (in particular the use of
$\lambda$-abstraction and $\beta\eta$-conversion) allows us to replace
mapping $\lfloor . \rfloor$ by mapping $|.|$ which works locally and is not
recursive.\footnote{Note that the encoding of the modal operators
  $\mball{r}$ is chosen to explicitly depend on an accessibility
  relation $r$ of type ${\iota\rightarrow\iota\rightarrow o}$ given as
  first argument to it.  Hence, we basically introduce a generic
  framework for modeling multimodal logics.  This idea is due to Brown
  and it is this aspect where the encoding differs from the LTL
  encoding of Harrison.  The latter chooses the interpreted type $num$
  of numerals and then uses the predefined relation $\leq$ over
  numerals as fixed accessibility relation in the definitions of
  $\mball{}$ and $\Diamond$.  By making the dependency of $\mball{r}$
  and $\Diamond_{r}$ on the accessibility relation $r$ explicit, we
  cannot only formalize but also automatically prove some meta
  properties of multimodal logics as we have demonstrated in
  \cite{B9}.}

It is easy to check that this local mapping works as
intended. For example, 
$$| \mball{r} p \mor\mball{r} q) | :=   |\mor|\,(|\mball{}|\,|r|\,|p|)\,(|\mball{}|\,|r|\,|q|) \eqbe \lfloor \mball{r} p \mor \mball{r} q) \rfloor$$

% $| s \mor t | := |s|\, |\mor|\, |t| \, \eqbe \lfloor s
% \mor t \rfloor$ and $ | \mball{r} s | := |mball{}| \, |r| \, |s| \eqbe \lfloor
% \mball{r} s \rfloor$.

Further local definitions for other multimodal logic operators can be introduced
this way. For example, $|\mimpl| = \lam{A_{\iota\rightarrow o}}
\lam{B_{\iota\rightarrow o}} \lam{X_\iota} (A\,X) \Rightarrow (B\,X)$,
$|\mfalse| = \lam{A_{\iota\rightarrow o}} \bot$, $|\mtrue| =
\lam{A_{\iota\rightarrow o}} \top$, and $|\mand| =
\lam{A_{\iota\rightarrow o}} \lam{B_{\iota\rightarrow o}}
\lam{X_\iota} (A\,X) \wedge (B\,X)$.
% Abbreviation based mappings like
% these can be very elegantly introduced as simple definitions in the
% higher-order theorem prover LEO-II; see Appendix \ref{app1} for the
% encoding of this mapping in the new TPTP THF syntax\footnote{Say
%   something}, which is also the input syntax of LEO-II.  LEO-II's
% termindexing support then guarantees efficient expansion of these
% definitions for proof automation.

A notion of validity for the $\lambda$-terms (of type $\iota\rightarrow
o$) we obtain after definition expansion is still missing: We want
$A_{\iota\rightarrow o}$ to be valid if and only if for all possible
worlds $w_\iota$ we have $(A_{\iota\rightarrow o}\,w_\iota)$, that is, $w\in A$.
%% Thus validity corresponds to the definition
%% of $\in$ in simple type theory.
This notion of validity is also introduced as a local definition:
\begin{eqnarray*}
  | \texttt{Mval} | & := & \lam{A_{\iota\rightarrow o}} \all{W_{\iota}} A\,W 
% satisfiable & := & \lam{A_{\iota\rightarrow o}} \exi{W_{\iota}} A\,W \\
% countersatisfiable & := & \lam{A_{\iota\rightarrow o}} \exi{W_{\iota}} \neg A\,W \\
% invalid & := & \lam{A_{\iota\rightarrow o}} \all{W_{\iota}} \neg A\,W \\
\end{eqnarray*}

Garg and Abadi's translation of access control into modal logic as
presented in Section \ref{gargabadi} is monomodal and does not require
different $\mball{r}$-operators.  Thus, for the purpose of this
paper we restrict the outlined general embedding of multimodal logics
into \stt to an embedding of monomodal logic into \stt. Hence, for the remainder of the paper we assume that
$\lm$ provides exactly one $\mball{r}$-operator, that is, a single
 relation constant $r$.
% Moerover, we assume that the in all Kripke frames the accessibility
% relation $R_r$ is reflexive and transitive.  This leads to the
% notion of S4-validity as we have introduced in Section
% \ref{preliminaries}: it

We next study soundness of this embedding. Our soundness proof below
employs the following mapping of Kripke frames into Henkin models.

\begin{definition}[Henkin model $M^K$ for Kripke Model
  $K$] \label{def} Given a Kripke model $K=\langle W,(R_r), \linebreak
  \models \rangle$. The Henkin model $M^K=\langle
  \{D_\alpha\}_{\alpha\in{\cal T}}, I \rangle$ for $K$ is defined as
  follows: We choose the set of individuals $D_\iota$ as the set of
  worlds $W$ and we choose the $D_{\alpha\rightarrow\beta}$ as the set
  of all functions from $D_\alpha$ to $D_\beta$. Let $p^1,\ldots,p^m$
  for $m\geq 1$ be the atomic primitives occuring in modal language
  $\lm$. Remember that $\mball{r}$ is the only box operator of $\lm$.
  Note that $| p^j | = p^j_{\iota\rightarrow o}$ and $| r | =
  r_{\iota\rightarrow\iota\rightarrow o}$.  Thus, for $1\leq i\leq m$
  we choose $I p^j_{\iota\rightarrow o} \in D_{\iota\rightarrow o}$
  such that $(I p^j_{\iota\rightarrow o})(w) = T$ for all $w\in
  D_\iota$ with $w\models p^j$ in Kripke model $K$ and $(I
  p^j_{\iota\rightarrow o})(w) = F$ otherwise. Similarly, we choose $I
  r_{\iota\rightarrow\iota\rightarrow o}\in
  D_{\iota\rightarrow\iota\rightarrow o}$ such that $(I
  r_{\iota\rightarrow\iota\rightarrow o})(w,w') = T$ if $R_{r}(w,w')$
  in Kripke model $K$ and $(I r_{\iota\rightarrow\iota\rightarrow
    o})(w,w') = F$ otherwise. Clearly, if $R_r$ is reflexive and
  transitive then, by construction, $I
  r_{\iota\rightarrow\iota\rightarrow o}$ is so as well.
% The denotation
%  of any constant different from $r_{\iota\rightarrow\iota o}$ and the
%  $p^j_{\iota\rightarrow o}$ can be chosen arbitrarily. Henkin
It is easy to check that
  $M^K=\langle \{D_\alpha\}_{\alpha\in{\cal T}}, I \rangle$ is a
  Henkin model. In fact it is a standard model since the function
  spaces are full.
%  In case the Kripke model $K$ is reflexive and
%   transitive a Henkin model $M^K$ can be constructed in which all
%   relations $r\in D_{\iota\rightarrow\iota\rightarrow o}$ are
%   reflexive and transitive.
\end{definition}

\begin{lemma} \label{lemma1} Let $M^K=\langle
  \{D_\alpha\}_{\alpha\in{\cal T}}, I \rangle$ be a Henkin model for
  Kripke model $K=\langle W,(R_i)_{i\in I},\models\rangle$.  For all $q\in L$,
  $w\in W$ and variable assignments $\phi$ the following are equivalent: 
 (i) $w \models q$, (ii) ${\cal V}_{[w/W_\iota],\phi}\, (\lfloor q \rfloor\, W) = T$, and (iii) ${\cal V}_{[w/W_\iota],\phi}\, (| q |\, W) = T$.
\end{lemma}

\begin{proof} We prove (i) if and only if (ii) by induction on the structure of $q$. Let $q=p$ for
  some atomic primitive $p\in L$. By construction of $M^K$, we have
  ${\cal V}_{[w/W_i],\phi}\, (\lfloor p \rfloor W) = {\cal
    V}_{[w/W_i],\phi}\, (p_{\iota\rightarrow o}\, W) =
  (I\,p_{\iota\rightarrow o})(w) = T$ if and only if $w \models
  p$. Let $p=\neg s$. We have $w \models \neg s$ if and only $w
  \not\models s$. By induction we get ${\cal V}_{[w/W_i],\phi}\,
  (\lfloor s \rfloor W) = F$ and hence ${\cal V}_{[w/W_i],\phi}\, \neg
  (\lfloor s \rfloor W) =_{\beta\eta} {\cal V}_{[w/W_i],\phi}\, (\lfloor \neg s
  \rfloor W) = T$.  Let $p=(s\vee t)$. We have $w \not\models (s\vee
  t)$ if and only if $w \models s$ or $w \not\models t$. By induction,
  ${\cal V}_{[w/W_\iota],\phi}\, (\lfloor s \rfloor\, W) = T$ or
  ${\cal V}_{[w/W_\iota],\phi}\, (\lfloor t \rfloor\, W) = T$. Thus
  ${\cal V}_{[w/W_\iota],\phi}\, (\lfloor s \rfloor\, W) \vee (\lfloor
  t \rfloor\, W) =_{\beta\eta} {\cal V}_{[w/W_\iota],\phi}\, (\lfloor s \vee t
  \rfloor\, W) = T$.  Let $q=\mball{r} s$. We have $w \models
  \mball{r} s$ if and only if for all $u$ with $R_r(w,u)$ we have
  $u\models s$.  By induction, for all $u$ with $R_r(w,u)$ we have
  ${\cal V}_{[u/V_\iota],\phi}\, (\lfloor s \rfloor\, V) = T$. Hence,
  ${\cal V}_{[u/V_\iota],[w/W_\iota],\phi}\, ((\lfloor r \rfloor\, W\,
  V) \Rightarrow (\lfloor s \rfloor\, V)) = T$ and thus ${\cal
    V}_{[w/W_\iota],\phi}\, (\all{Y_\iota} ((\lfloor r \rfloor\, W\, Y)
  \Rightarrow (\lfloor s \rfloor\, Y))) =_{\beta\eta} {\cal V}_{[w/W_\iota],\phi}\,
  (\lfloor \mball{r} s \rfloor\, W) = T$. 

  We leave it to the reader to prove (ii) if and only if (iii).
\end{proof}

% \begin{lemma} Given a Kripke frame $K=\langle
%   W,(R_i)_{i\in I}\rangle$ and a Henkin model $M^K=\langle
%   \{D_\alpha\}_{\alpha\in{\cal T}}, I \rangle$ for $K$ as constructed
%   above. 
% \begin{enumerate}
% \item \label{aa} If $K$ is reflexive then there exists 
%   $(\all{R_{\iota\rightarrow\iota\rightarrow
%       o}}\all{X_{\iota\rightarrow o}} \texttt{valid}\, \lfloor
%   \mball{R} X \mimpl X\rfloor)$ is valid in $M^K$.
% \item \label{bb} 
% \end{enumerate}
% \end{lemma}
% \begin{proof}
%   (\ref{aa}) $(\all{R_{\iota\rightarrow\iota\rightarrow
%       o}}\all{X_{\iota\rightarrow o}} \texttt{valid}\, \lfloor
%   \mball{R} X \mimpl X\rfloor)$ is valid in $M^K$ if and only
%   if ${\cal V}_{\phi}\, (\all{R_{\iota\rightarrow\iota\rightarrow
%       o}}\all{X_{\iota\rightarrow o}} \all{W_\iota} ((\all{Y_\iota}
%   (R\,W\,Y) \Rightarrow (X\,Y)) \Rightarrow (X\,W)))$.

% ${\cal V}_{[w/W_i],\phi}\,$

%   $(\all{R_{\iota\rightarrow\iota\rightarrow
%       o}}\all{X_{\iota\rightarrow o}} \texttt{valid}\, \lfloor
%   \mball{R} X \mimpl X\rfloor)$ 
% \end{proof}

\pagebreak

We now prove soundness of the embedding of normal monomodal logics $K$
and $S4$ into \stt. In the case of $S4$ we add axioms
that correspond to modal logic axioms $T$ (reflexivity) and $4$
(transitivity).\footnote{Note that $T=(\mball{r} s \mimpl s)$ and
  $4=(\mball{r} s \mimpl \mball{r} \mball{r} s)$ are actually axiom
  schemata in modal logic. As we show here, their counterparts in
  \stt actually become proper axioms.} Here we call these axiom \texttt{R} and \texttt{T}.
\begin{theorem}[Soundness of the Embedding of $K$ and $S4$ into $\stt$] \label{thm1}
  Let $s\in \lm$ be a mono\-modal logic proposition.
  % and let $\lfloor A \rfloor \in L^{\ttremoveclash}$ be its corresponding term in
  % simple type theory.
\begin{enumerate}
\item \label{a} If $\models^{\ttremoveclash} | \texttt{Mval}\,\, s |$ then $\models^{K}s$.
\item \label{b} If $\{\texttt{R,T}\}\models^{\ttremoveclash} | \texttt{Mval}\,\, s|$
  then $\models^{S4}s$, where \texttt{R} and \texttt{T} are
  shorthands for \linebreak $\all{X_{\iota\rightarrow o}} |
  \texttt{Mval}\,\,\mball{r} X \mimpl X|$ and $\all{X_{\iota\rightarrow
      o}} |\texttt{Mval}\,\, \mball{r} X \mimpl \mball{r} \mball{r} X|$
  respectively.
\end{enumerate}
\end{theorem}

\begin{proof}
  
  (\ref{a}) The proof is by contraposition. For this, assume
  $\not\models^{K}s$, that is, there is a Kripke model $K=\langle
  W,(R_r),\models\rangle$ with $w \not\models s$ for some $w\in
  W$. By Lemma \ref{lemma1}, for arbitrary $\phi$ we have ${\cal
    V}_{[w/W_\iota],\phi}\, (| s |\, W) = F$ in Henkin
  model $M^K$ for $K$.  Thus, ${\cal V}_{\phi}\, (\all{W_\iota}
  (| s |\, W) = {\cal V}_{\phi}\, |\texttt{Mval}\, 
  s | = F$. Hence, $\not\models^{\ttremoveclash} |\texttt{Mval}\, s|$. 
  
  (\ref{b}) The proof is by contraposition. From
  $\not\models^{S4}s$ we get by Lemma \ref{lemma1} that
  $|\texttt{Mval}\, s |$ is not valid in Henkin model $M^K=\langle
  \{D_\alpha\}_{\alpha\in{\cal T}}, I \rangle$ for Kripke model
  $K=\langle W,(R_r)\rangle$. $R_r$ in $K$ is reflexive and
  transitive, hence, the relation $(I r) \in D_{\iota\rightarrow\iota
    \rightarrow o}$ is so as well. We leave it to the
  reader to verify that axioms \texttt{R} and \texttt{T} are valid in
  $M^K$. Hence, $\{\texttt{R,T}\}\not\models^{\ttremoveclash} |\texttt{Mval}\,
  s|$.
\end{proof}

In order to prove completeness, we introduce a mapping from Henkin
models to Kripke models.  We assume that the the signature of the
modal logic contains the atomic primitives $p^1,\ldots,p^m$ for $m\geq
1$ and that the simple type theory signature correspondingly contains
constants $p^1_{\iota\rightarrow o},\ldots,p^m_{\iota\rightarrow o}$
for $m\geq 1$ as well as relation constant $r_{\iota\rightarrow\iota\rightarrow o}$.

\begin{definition}[Kripke Model $K^M$ for Henkin model
  $M$] \label{def2} Let Henkin model $M = \langle
  \{D_\alpha\}_{\alpha\in{\cal T}}, \linebreak I \rangle$ be given. The Kripke model
  $K^M=\langle W,(R_r), \models \rangle$ for Henkin model $M$ is
  defined as follows: We choose the set of worlds $W$ as the set of
  individuals $D_\iota$. Moreover, we choose $\models$ such that
  $w\models p^i$ in $K^M$ if $(I p^j_{\iota\rightarrow o})(w) = T$ in
  $M$ and $w\not\models p^i$ otherwise. Similarly, we choose $R_r$
  such that $w\,R_r\,w'$ in $K^M$ if $(I
  r_{\iota\rightarrow\iota\rightarrow o})(w,w') = T$ in $M$ and
  $\neg(w\,R_r\,w')$ otherwise. Clearly , if $(I
  r_{\iota\rightarrow\iota\rightarrow o})$ is reflexive and transitive
  then also $R_r$ is. It is easy to check that $K^M$ is a Kripke
  model.
 \end{definition}

 \begin{lemma} \label{lemma2} Let $K^M=\langle W,(R_i)_{i\in
     I},\models\rangle$ be a Kripke model for Henkin model $M=\langle
   \{D_\alpha\}_{\alpha\in{\cal T}}, I \rangle$.  For all $q\in L$,
   $w\in W$ and variable assignments $\phi$ the following are
   equivalent: (i) $w \models q$, (ii) ${\cal V}_{[w/W_\iota],\phi}\,
   (\lfloor q \rfloor\, W) = T$, and (iii) ${\cal
     V}_{[w/W_\iota],\phi}\, (| q |\, W) = T$.
\end{lemma}

\begin{proof} 
Analogous to Lemma \ref{lemma1}.
\end{proof}

\pagebreak

We now prove completeness of the embedding of normal monomodal logics $K$
and $S4$ into \stt. As before we add axioms $T$ and $R$ to obtain $S4$.
\begin{theorem}[Completeness of the Embedding of $K$ and $S4$ into $\stt$] \label{thm2}
  Let $s\in \lm$ be a mono\-modal logic proposition.
  % and let $\lfloor A \rfloor \in L^{\ttremoveclash}$ be its corresponding term in
  % simple type theory.
\begin{enumerate}
\item \label{aa} If  $\models^{K}s$ then $\models^{\ttremoveclash} | \texttt{Mval}\,\, s |$.
\item \label{bb} If $\models^{S4}s$ then $\{\texttt{R,T}\}\models^{\ttremoveclash} | \texttt{Mval}\,\, s|$, where \texttt{R} and \texttt{T} are
  shorthands for \linebreak $\all{X_{\iota\rightarrow o}} |
  \texttt{Mval}\,\,\mball{r} X \mimpl X|$ and $\all{X_{\iota\rightarrow
      o}} |\texttt{Mval}\,\, \mball{r} X \mimpl \mball{r} \mball{r} X|$
  respectively.
\end{enumerate}
\end{theorem}

\begin{proof}
  
  (\ref{a}) The proof is by contraposition. Assume $\not\models^{\ttremoveclash}
  | \texttt{Mval}\,\, s |$, that is, for a Henkin model $M=\langle
  \{D_\alpha\}_{\alpha\in{\cal T}}, I \rangle$ and a variable
  assignment $\phi$ we have ${\cal V}_{\phi}\, | \texttt{Mval}\,\, s |
  = F$ in $M$. This implies that there is some $w\in D_i$ such that
  ${\cal V}_{[w/W_\iota],\phi}\, (| s |\, W) = F$ in $M$. By Lemma
  \ref{lemma2} we know that $w \not\models s$ in Kripke model
  $K^M=\langle W,(R_r),\models\rangle$ for $M$. Hence,
  $\not\models^{K}s$.
  
  (\ref{b}) The proof is analogous to above and from
  $\{\texttt{R,T}\}\not\models^{\ttremoveclash} | \texttt{Mval}\,\, s|$ we get
  with Lemma \ref{lemma2} that $w \not\models s$ in Kripke model
  $K^M=\langle W,(R_r),\models\rangle$ for $M$. However, we now
  additionally have for axioms \texttt{R} and \texttt{T} that ${\cal
    V}_{\phi}\,\texttt{R} = {\cal V}_{\phi}\,\texttt{T} = T$. We leave
  it to the reader to check that this implies reflexivity and
  transitivity of relation $(I r_{\iota\rightarrow\iota\rightarrow
    o})$. Thus, by construction, $R_r$ in $K^M$ is reflexive and
  transitive. This implies $\not\models^{S4}s$.
\end{proof}

Reasoning problems in modal logics $K$ and $S4$ can thus be considered
as reasoning problems in \stt. Hence, any off the shelf theorem prover
that is sound for \stt, such as our LEO-II, can be applied to them.
For example, $\models^{\ttremoveclash} | \texttt{Mval}\, \mball{r}\,\mtrue|$,
$\models^{\ttremoveclash} |\texttt{Mval}\, \mball{r}a \mimpl \mball{r}a|$, and
$\models^{\ttremoveclash} | \texttt{Mval}\, \Diamond_{r}(a \mimpl
b)\mor(\mball{r}a \mimpl \mball{r}b)|$ are automatically proved by
LEO-II in 0.024 seconds, 0.026 seconds, and 0.035 seconds
respectively. All experiments with LEO-II reported in this paper
were conducted with LEO-II version v0.98 \footnote{LEO-II is available
  from \url{http://www.ags.uni-sb.de/~leo/}.} on a notebook computer
with a Intel Pentium 1.60GHz processor with 1GB memory running Linux.

More impressive example problems illustrating LEO-II's performance for
reasoning in and \textit{about} multimodal logic can be found in
\cite{B9}. Amongst these problems is also the equivalence between
axioms $\mball{r} s \mimpl s$ and $\mball{r} s \mimpl \mball{r}
\mball{r} s$ and the reflexivity and transitivity properties of the
accessibility relation $r$:
\begin{example} \label{ex4} $\models^{\ttremoveclash} (\texttt{R} \wedge
  \texttt{T}) \Leftrightarrow (\texttt{refl}\,r \wedge
  \texttt{trans}\,r)$ where \texttt{R} and \texttt{T} are the
  abbreviations as introduced in Theorem \ref{thm1} and \texttt{refl}
  and \texttt{trans} abbreviations for
  $\lam{R_{\iota\rightarrow\iota\rightarrow o}} \all{X_\iota} R\,X\,X$
  and $\lam{R_{\iota\rightarrow\iota\rightarrow o}} \all{X_\iota}
  \all{Y_\iota} \all{Z_\iota} R\,X\,Y \wedge R\,Y\,Z \Rightarrow
  R\,X\,Z$. LEO-II can solve this modal logic meta-level problem in
  2.329 seconds.
\end{example}

\vfill\pagebreak

\section{Embedding Access Control Logic in Simple Type
  Theory} \label{combined} We combine the results from Sections
\ref{gargabadi} and \ref{benzpaulson} and obtain the following 
mapping $\|.\|$ from access control logic $ICL$ into \stt:
\begin{eqnarray*}
  \|p\| & = & | \mball{r} p | = \lam{X_\iota} \all{Y_\iota} r_{\iota\rightarrow\iota\rightarrow o}\, X\, Y \Rightarrow p_{\iota\rightarrow o}\, Y \\
  \|A\| &  = & | A | = a_{\iota\rightarrow o} \,\,\,\, \texttt{(distinct from the $p_{\iota\rightarrow o}$)}\\   
  \|\mand\| & = &  \lam{S} \lam{T} | S \mand T | = \lam{S_{\iota\rightarrow o}} \lam{T_{\iota\rightarrow o}} \lam{X_\iota} S\,X \wedge T\,X \\
  \|\mor\| & = &   \lam{S} \lam{T} | S \mor T | = \lam{S_{\iota\rightarrow o}} \lam{T_{\iota\rightarrow o}} \lam{X_\iota} S\,X \vee T\,X \\
  \|\supset\| & = &  \lam{S} \lam{T} | \mball{r} (S \mimpl T) | \\
 & = & \lam{S_{\iota\rightarrow o}} \lam{T_{\iota\rightarrow o}} \lam{X_\iota} 
\all{Y_\iota}  r_{\iota\rightarrow\iota\rightarrow o}\, X\, Y \Rightarrow (S\,Y \Rightarrow T\,Y) \\
  \|\mtrue\| & = & | \mtrue | = \lam{S_{\iota\rightarrow o}} \top \\
  \|\mfalse\| & = & | \mfalse | =  \lam{S_{\iota\rightarrow o}} \bot \\
  \|\texttt{says}\| & = &   \lam{A} \lam{S} | \mball{r}(A \vee S) | \\
 & = & \lam{A_{\iota\rightarrow o}}  \lam{S_{\iota\rightarrow o}}  \lam{X_\iota} \all{Y_\iota} r_{\iota\rightarrow\iota\rightarrow o}\, X\, Y \Rightarrow (A\, Y \vee S\, Y)
\end{eqnarray*}
It is easy to verify that this mapping works as
intended. For example:
\begin{eqnarray*}
\| \texttt{admin}\,\,\texttt{says}\,\,\mfalse\| & := & \|\texttt{says}\| \|\texttt{admin}\| \|\mfalse\| \\
& \eqbe & \lam{X_\iota} \all{Y_\iota} r_{\iota\rightarrow\iota\rightarrow o}\, X\, Y \Rightarrow (\texttt{admin}_{\iota\rightarrow o}\, Y \vee \bot) \\
 & \eqbe & | \mball{r} (\texttt{admin}\, \mor\, \mfalse) | \eqbe  \lfloor \mball{r} (\texttt{admin}\, \mor\, \mfalse) \rfloor \\
 & = & \lfloor \lceil \texttt{admin}\,\,\texttt{says}\,\, \mfalse \rceil \rfloor 
\end{eqnarray*}

We extend this mapping to logic $ICL^{\Rightarrow}$ by adding a clause for the speaks-for connective $\iclimplprinc$:
$$ \| \iclimplprinc \|   =  \lam{A} \lam{B} | \mball{r} (A \mimpl B) | = \lam{A_{\iota\rightarrow o}}  \lam{B_{\iota\rightarrow o}}  \lam{X_\iota} \all{Y_\iota} r_{\iota\rightarrow\iota\rightarrow o}\, X\, Y \Rightarrow (A\, Y \Rightarrow B\, Y) $$

For the translation of $ICL^{B}$ we simply allow that the ICL
connectives can be applied to principals. Our mapping $\|.\|$ needs not
to be modified and is applicable as is.

The notion of validity for the terms we obtain after translations is 
chosen identical to before
$$
  \|\texttt{ICLval} \| = \lam{A_{\iota\rightarrow o}} | \texttt{Mval} | \, A = \lam{A_{\iota\rightarrow o}} \all{W_{\iota}} A\,W  
$$

\begin{theorem}[Soundness of the Embeddings of $ICL$, $ICL^{\Rightarrow}$, and $ICL^{B}$ in $\stt$]
  Let $s\in ICL$ (resp. $s\in ICL^{\Rightarrow}$, $s\in ICL^{B}$) 
  and let \texttt{R} and \texttt{T} be as before.
% for
%   $\all{X_{\iota\rightarrow o}} \texttt{Mval}\, \lfloor
%   \mball{r} X \mimpl X\rfloor$ and
%   $\all{X_{\iota\rightarrow o}} \texttt{Mval}\, \lfloor
%   \mball{r} X \mimpl \mball{r} \mball{r} X \rfloor$ respectively..
  If $\{\texttt{R,T}\}\models^{\ttremoveclash} |\texttt{ICLval}\, s|$ then
  $\vdash s$ in access control logic $ICL$ (resp. $ICL^{\Rightarrow}$,
  $ICL^{B}$).
\end{theorem}
\begin{proof} If $\{\texttt{R,T}\}\models^{\ttremoveclash} |\texttt{ICLval}\, s
  |$ then $\models^{S4}s$ by Theorem \ref{thm1} since
  $|\texttt{ICLval}\, s | = |\texttt{Mval}\, s |$. This implies that
  $\vdash \lceil s \rceil$ for the sound and complete Hilbert System
  for S4 studied in \cite{GargAbadi08}.\footnote{See Theorem 8 in
    \cite{GargAbadi08} which is only given in the full version of the
    paper available from
    \url{http://www.cs.cmu.edu/~dg/publications.html}.} By Theorem
  \ref{thm:gargabadi} we conclude that $\vdash s$ in access control
  logic $ICL$ (resp. $ICL^{\Rightarrow}$, $ICL^{B}$).
\end{proof}

\pagebreak

\begin{theorem}[Completeness of the Embeddings of $ICL$, $ICL^{\Rightarrow}$, and $ICL^{B}$ in $\stt$]
  Let $s\in ICL$ (resp. $s\in ICL^{\Rightarrow}$, $s\in ICL^{B}$) 
  and let \texttt{R} and \texttt{T} be as before.
% for
%   $\all{X_{\iota\rightarrow o}} \texttt{Mval}\, \lfloor
%   \mball{r} X \mimpl X\rfloor$ and
%   $\all{X_{\iota\rightarrow o}} \texttt{Mval}\, \lfloor
%   \mball{r} X \mimpl \mball{r} \mball{r} X \rfloor$ respectively..
  If $\vdash s$ in access control logic $ICL$
  (resp. $ICL^{\Rightarrow}$, $ICL^{B}$) then
  $\{\texttt{R,T}\}\models^{\ttremoveclash} |\texttt{ICLval}\, s|$
\end{theorem}
\begin{proof} Similar to above with  Theorem \ref{thm2} instead of Theorem \ref{thm1}.
\end{proof}

\begin{table} \caption{Performance of LEO-II when applied to problems in access control logic $ICL$} \label{table1}
\centering \footnotesize
\begin{tabular}{|lllr|} \hline
 {\bf Name} &  {\bf TPTP Name} & {\bf Problem} & {\bf LEO (s)} \\ \hline 
& & \\[-.95em] 
unit & \url{SWV425^1.p} & $\{\texttt{R,T}\}\models^{\ttremoveclash}  \| \texttt{ICLval}\,\, s \iclimpl (A \iclsays s) \| $  &  0.031 \\
cuc & \url{SWV426^1.p} & $\{\texttt{R,T}\}\models^{\ttremoveclash} \| \texttt{ICLval}\,\, (A \iclsays (s \iclimpl t)) \iclimpl (A \iclsays s) \iclimpl (A \iclsays t) \|$ & 0.083 \\
idem & \url{SWV427^1.p} & $\{\texttt{R,T}\}\models^{\ttremoveclash} \| \texttt{ICLval}\,\,  (A \iclsays A \iclsays s) \iclimpl (A \iclsays s) \|$ & 0.037 \\
Ex1 & \url{SWV428^1.p} & $\{\texttt{R},\texttt{T},\|\texttt{ICLval}\,\, (1.1)\|,\ldots,\| \texttt{ICLval}\,\, (1.3)\|\}\models^{\ttremoveclash} \| \texttt{ICLval}\,\, (1.4)\|$ &  3.494 \\ \hline
& & \\[-.95em] 
unit$^K$ & \url{SWV425^2.p} & $\models^{\ttremoveclash} \| \texttt{ICLval}\,\, s \iclimpl (A \iclsays s) \| $  &  -- \\
cuc$^K$ & \url{SWV426^2.p} & $ \models^{\ttremoveclash} \| \texttt{ICLval}\,\, (A \iclsays (s \iclimpl t)) \iclimpl (A \iclsays s) \iclimpl (A \iclsays t) \|$ & -- \\
idem$^K$ & \url{SWV427^2.p} &  $ \models^{\ttremoveclash} \| \texttt{ICLval}\,\, (A \iclsays A \iclsays s) \iclimpl (A \iclsays s) \|$ & -- \\ 
Ex1$^K$ & \url{SWV428^2.p} & $\{\| \texttt{ICLval}\,\, (1.1)\|,\ldots,\| \texttt{ICLval}\,\, (1.3)\|\}\models^{\ttremoveclash} \| \texttt{ICLval}\,\, (1.4)\|$ & -- \\
\hline
\end{tabular}
\end{table}

\begin{table} \caption{Performance of LEO-II when applied to problems in access control logic $ICL^\Rightarrow$} \label{table2}
\centering \footnotesize
\begin{tabular}{|lllr|} \hline
 {\bf Name} &  {\bf TPTP Name} & {\bf Problem} & {\bf LEO (s)} \\ \hline 
& & \\[-.95em] 
refl  & \url{SWV429^1.p}       & $\{\texttt{R,T}\}\models^{\ttremoveclash} \| \texttt{ICLval}\,\,  A \iclimplprinc A \| $  & 0.052 \\
trans & \url{SWV430^1.p}       & $\{\texttt{R,T}\}\models^{\ttremoveclash} \| \texttt{ICLval}\,\, (A \iclimplprinc B) \iclimpl (B \iclimplprinc C) \iclimpl (A \iclimplprinc C) \|$   &  0.105 \\
sp.-for & \url{SWV431^1.p}     & $\{\texttt{R,T}\}\models^{\ttremoveclash} \| \texttt{ICLval}\,\, (A \iclimplprinc B) \iclimpl (A \iclsays s) \iclimpl (B \iclsays s) \|$   &   0.062 \\
handoff & \url{SWV432^1.p}     & $\{\texttt{R,T}\}\models^{\ttremoveclash} \| \texttt{ICLval}\,\, (B \iclsays (A \iclimplprinc B)) \iclimpl (A \iclimplprinc B) \| $  &   0.036 \\
Ex2    & \url{SWV433^1.p}      & $\{\texttt{R},\texttt{T},\| \texttt{ICLval}\,\, (2.1)\|,\ldots,\| \texttt{ICLval}\,\, (2.4)\|\}\models^{\ttremoveclash} \| \texttt{ICLval}\,\, (2.5)\|$ &   0.698 \\ \hline 
& & \\[-.95em] 
refl$^K$ & \url{SWV429^2.p}    & $\models^{\ttremoveclash} \| \texttt{ICLval}\,\, A \iclimplprinc A \| $  & 0.031  \\
trans$^K$ & \url{SWV430^2.p}   & $\models^{\ttremoveclash} \| \texttt{ICLval}\,\, (A \iclimplprinc B) \iclimpl (B \iclimplprinc C) \iclimpl (A \iclimplprinc C) \|$   & -- \\
sp.-for$^K$ & \url{SWV431^2.p}  & $\models^{\ttremoveclash} \| \texttt{ICLval}\,\, (A \iclimplprinc B) \iclimpl (A \iclsays s) \iclimpl (B \iclsays s) \|$   & --  \\
handoff$^K$ & \url{SWV432^2.p}  & $\models^{\ttremoveclash} \| \texttt{ICLval}\,\, (B \iclsays (A \iclimplprinc B)) \iclimpl (A \iclimplprinc B) \| $  &  -- \\
Ex2$^K$  & \url{SWV433^2.p}    & $\{\| \texttt{ICLval}\,\, (2.1)\|,\ldots,\| \texttt{ICLval}\,\, (2.4)\|\}\models^{\ttremoveclash} \| \texttt{ICLval}\,\, (2.5)\|$ &  -- \\
\hline
\end{tabular}
\end{table}

\begin{table}[t] \caption{Performance of LEO-II when applied to problems in access control logic $ICL^B$} \label{table3}
\centering \footnotesize
\begin{tabular}{|lllr|} \hline
 {\bf Name} &  {\bf TPTP Name} & {\bf Problem} & {\bf LEO (s)} \\ \hline 
& & \\[-.95em] 
trust  & \url{SWV434^1.p}      & $\{\texttt{R,T}\}\models^{\ttremoveclash} \| \texttt{ICLval}\,\, (\iclfalse \iclsays s) \iclimpl s\| $  &  0.049 \\
untrust  & \url{SWV435^1.p}    & $\{\texttt{R,T},\| \texttt{ICLval}\,\, A\equiv \icltrue\|\}\models^{\ttremoveclash}  \| \texttt{ICLval}\,\, A \iclsays \iclfalse\|$   &  0.053 \\
cuc'   & \url{SWV436^1.p}      & $\{\texttt{R,T}\}\models^{\ttremoveclash}  \| \texttt{ICLval}\,\, ((A \iclimpl B) \iclsays s) \iclimpl  (A \iclsays s) \iclimpl (B \iclsays s) \|$   &   0.131 \\
Ex3   & \url{SWV437^1.p}       & $\{\texttt{R},\texttt{T}, \| \texttt{ICLval}\,\, (3.1)\|,\ldots, \| \texttt{ICLval}\,\, (3.3)\|\}\models^{\ttremoveclash}  \| \texttt{ICLval}\,\, (3.4)\|$ &   0.076 \\ \hline 
& & \\[-.95em] 
trust$^K$  & \url{SWV434^2.p}  & $\models^{\ttremoveclash}  \| \texttt{ICLval}\,\, (\iclfalse \iclsays s) \iclimpl s \| $  & -- \\
untrust$^K$ & \url{SWV435^2.p}  & $\{\| \texttt{ICLval}\,\, A\equiv \icltrue\|\} \models^{\ttremoveclash}  \| \texttt{ICLval}\,\,  A \iclsays \iclfalse\|$   & 0.041 \\
cuc'$^K$  & \url{SWV436^2.p}   & $\models^{\ttremoveclash}  \| \texttt{ICLval}\,\, ((A \iclimpl B) \iclsays s) \iclimpl  (A \iclsays s) \iclimpl (B \iclsays s) \|$   & -- \\
Ex3$^K$   & \url{SWV437^2.p}   & $\{ \| \texttt{ICLval}\,\, (3.1)\|,\ldots, \| \texttt{ICLval}\,\, (3.3)\|\}\models^{\ttremoveclash}  \| \texttt{ICLval}\,\, (3.4)\|$ &  --  \\
\hline
\end{tabular}
\end{table}

We can thus safely exploit our framework to map problems formulated in
the control logics $ICL$, $ICL^{\Rightarrow}$, and $ICL^{B}$ to
problems in \stt and we can apply the off the shelf higher-order
theorem prover LEO-II (which itself cooperates with the first-order theorem
prover E \cite{Schulz:AICOM-2002}) to solve them. Times are
given in seconds.

Table \ref{table1} shows that LEO-II can effectively prove that the
axioms unit, cuc and idem hold as expected in our embedding of $ICL$
in \stt. This provides additional evidence for the correctness of our
approach. Example 1 can also be quickly solved by LEO-II. Problems
unit$^K$, cuc$^K$, idem$^K$, and Ex1$^K$ modify their counterparts by
omitting the axioms \texttt{R} and \texttt{T}. Thus, they essentially
test whether these problems can already be proven via a mapping to
modal logic $K$ instead of $S4$, which is not expected.  A challenge
for future work is to apply LEO-II to analyse invalidity of these
axioms in context $K$ and to synthesize concrete witness terms if
possible.  For unit$^K$, for instance, the problem given to LEO-II
would be
$$\models^{\ttremoveclash} \exi{s} \neg \| \texttt{ICLval}\,\, s \iclimpl (A \iclsays s) \| $$

Tables \ref{table2} and \ref{table3} extend our experiment to the
other access control logics, axioms and examples presented in Section
\ref{gargabadi}.  In the cases of refl$^K$ for logic
$ICL^{\Rightarrow}$ and untrust$^K$ for logic $ICL^{B}$ LEO-II shows
that the axioms \texttt{R} and \texttt{T} are in fact not needed.

In the Appendix we present the concrete encoding or our embedding
together with the problems unit, cuc, idem, and Ex1 in the new TPTP
THF syntax \cite{C25}, which is also the input syntax of LEO-II.

\section{Conclusion and Future Work} \label{conc} We have outlined a
framework for the automation of reasoning in and about different
access control logics in simple type theory. Using our framework off
the shelf higher-order theorem provers and proof assistants can be
applied for the purpose. Our embedding of access control logics in
simple type theory and a selection of example problems have been
encoded in the new TPTP THF syntax and our higher-order theorem prover
LEO-II has been applied to them yielding promising initial
results. Our problem encodings have been submitted to the higher-order
TPTP library (see \url{http://www.cs.miami.edu/~tptp/}; problem domain
\texttt{thf}) under development in the EU project THFTPTP and are thus
available for comparison and competition with other TPTP compliant
theorem provers.

Future work includes the evaluation of the scalability of our
approach for reasoning within prominent access control logics.
% It also motivates the study of the simple type theory target
% fragments of our embeddings (for example, do decidability results for
% the source logics hint at unexplored decision procedures within the
% simple type theory target fragments?).
Moreover, LEO-II could be applied to explore meta-properties of access
control logics $ICL$, $ICL^{\Rightarrow}$, and $ICL^{B}$ analogous to
Example \ref{ex4}. More generally, we would like to study whether our
framework can fruitfully support the exploration of new access control
logics.
% Moreover, higher-order theorem provers such as LEO-II typically
% generate proof objects so that the reasoning results can generally be
% independently verified.

What has not been addressed in this paper due to space restrictions is
our embedding of access control logic $ICL^\forall$ into simple type
theory -- $ICL^\forall$ is an access control logic with second-order
quantification.

\paragraph*{Acknowledgements:} Catalin Hritcu inspired the work
presented in this paper and pointed me to the paper by Garg and
Abadi. Chad Brown, Larry Paulson and Claus-Peter Wirth pointed me to
some problems and typos in earlier versions of this paper.

\vfill\pagebreak

\bibliographystyle{plain}
\bibliography{techreport}

\mbox{}

\section{TPTP THF Problem files for Ex1}
%\paragraph*{File ICL\_k.ax}
The file ICL\_k.ax presents the general definitions of our mapping 
from access control logics via modal logic K to \stt.
{\scriptsize
%\verbatiminput{ICL\_s4.ax}
\begin{verbatim}
%------------------------------------------------------------------------------
% File     : ICL_k.ax
% Domain   : ICL Logic and its translation into Modal Logic (which is
%            itself modeled in simple type theory; see [2])
% Axioms   : ICL logic based upon modal logic based upon simple type theory
% Version  : 
% English  :
% Refs     : [1] Deepak Garg, Martín Abadi: A Modal Deconstruction of Access 
%            Control Logics. FoSSaCS 2008: 216-230
%            [2] C. Benzmueller and L. Paulson. Exploring Properties 
%            of Normal Multimodal Logics in Simple Type Theory with LEO-II.
%            Festschrift in Honour of Peter B. Andrews. 
%            See: http://www.ags.uni-sb.de/~chris/papers/B9.pdf 
% Status   :
% Syntax   : 
% Comments : Formalization in THF by C. Benzmueller 
%------------------------------------------------------------------------------

%%%%%%%%%%%%%%%%%%%%%%%
%% Multimodal-Logic %%
%%%%%%%%%%%%%%%%%%%%%%%

%----This formalization of multimodal Logic follows the ideas presented in [2]

%----The idea is that an atomic multimodal logic proposition P (of type 
%---- $i >  $o) holds at a world W (of type  $i) iff W is in P resp. (P @ W)
%----Now we define the multimodal logic connectives by reducing them to set 
%----operations
%----mfalse corresponds to emptyset (of type $i)
thf(mfalse_decl,type,(
    mfalse: $i > $o )).

thf(mfalse,definition,
    ( mfalse
    := ( ^ [X: $i] : $false ) )).

%----mtrue corresponds to the universal set (of type $i)
thf(mtrue_decl,type,(
    mtrue: $i > $o )).

thf(mtrue,definition,
    ( mtrue
    := ( ^ [X: $i] : $true ) )).

%----mnot corresponds to set complement
thf(mnot_decl,type,(
    mnot: ( $i > $o ) > $i > $o )).

thf(mnot,definition,
    ( mnot
    := ( ^ [X: $i > $o,U: $i] : 
          ~ ( X @ U ) ) )).

%----mor corresponds to set union 
thf(mor_decl,type,(
    mor: ( $i > $o ) > ( $i > $o ) > $i > $o )).

thf(mor,definition,
    ( mor
    := ( ^ [X: $i > $o,Y: $i > $o,U: $i] : 
          ( ( X @ U )
          | ( Y @ U ) ) ) )).

%----mand corresponds to set intersection 
thf(mand_decl,type,(
    mand: ( $i > $o ) > ( $i > $o ) > $i > $o )).

thf(mand,definition,
    ( mand
    := ( ^ [X: $i > $o,Y: $i > $o,U: $i] : 
          ( ( X @ U )
          & ( Y @ U ) ) ) )).

%----mimpl defined via mnot and mor
thf(mimpl_decl,type,(
    mimpl: ( $i > $o ) > ( $i > $o ) > $i > $o )).


thf(mimpl,definition,
    ( mimpl
    := ( ^ [U: $i > $o,V: $i > $o] : 
          ( mor @ ( mnot @ U ) @ V ) ) )).

%----miff defined via mand and mimpl
thf(miff_decl,type,(
    miff: ( $i > $o ) > ( $i > $o ) > $i > $o )).

thf(miff,definition,
    ( miff
    := ( ^ [U: $i > $o,V: $i > $o] : 
          ( mand @ ( mimpl @ U @ V ) @ ( mimpl @ V @ U ) ) ) )).

%----mbox
thf(mbox_decl,type,(
    mbox: ( $i > $i > $o ) > ( $i > $o ) > $i > $o )).

thf(mbox,definition,
    ( mbox
    := ( ^ [R: $i > $i > $o,P: $i > $o,X: $i] : 
        ! [Y: $i] : 
          ( ( R @ X @ Y )
         => ( P @ Y ) ) ) )).

%----mdia
thf(mdia_decl,type,(
    mdia: ( $i > $i > $o ) > ( $i > $o ) > $i > $o )).

thf(mdia,definition,
    ( mdia
    := ( ^ [R: $i > $i > $o,P: $i > $o,X: $i] : 
        ? [Y: $i] : 
          ( ( R @ X @ Y )
          & ( P @ Y ) ) ) )).

%----Validity of a multimodal logic formula (in logic K) can now be encoded as
thf(mvalid_decl,type,(
    mvalid: ( $i > $o ) > $o )).

thf(mvalid,definition,
    ( mvalid
    := ( ^ [P: $i > $o] : 
        ! [W: $i] : 
          ( P @ W ) ) )).

%%%%%%%%%%%%%%%%%
%%% ICL Logic %%%
%%%%%%%%%%%%%%%%%

%----The encoding of ICL logic employs only one accessibility relation which
%----introduce here as a constant 'rel'; we don't need multimodal logic.
thf(rel,type,(
    rel : $i > $i > $o )).

%----ICL logic distiguishes between atoms and principals; for this we introduce
%----a predicate 'icl_atom' ...
thf(icl_atom,type,(
    icl_atom:  ($i > $o) > ($i > $o) )).

thf(icl_atom,definition,
    ( icl_atom
    := ( ^ [P: $i > $o] : (mbox @ rel @ P)) )).

%---- ... and also a predicate 'icl_princ'
thf(icl_princ,type,(
    icl_princ:  ($i > $o) > ($i > $o) )).

thf(icl_princ,definition,
    ( icl_princ
    := ( ^ [P: $i > $o] : P) )).

%%%%%%%%%%%%%%%%%%%%%%%%%%%%%%%%%%%%%%%%%%%%%%%%%%%%%%%
% We introduce the logical connectives of ICL and map %
% them to modal logic expressions as suggested in [1] %
%%%%%%%%%%%%%%%%%%%%%%%%%%%%%%%%%%%%%%%%%%%%%%%%%%%%%%%

%----ICL and connective
thf(icl_and,type,(
    icl_and:  ($i > $o) > ($i > $o) > ($i > $o) )).

thf(icl_and,definition,
    ( icl_and
    := ( ^ [A: $i > $o, B: $i > $o] : (mand @ A @ B)) )).

%----ICL or connective
thf(icl_or,type,(
    icl_or:  ($i > $o) > ($i > $o) > ($i > $o) )).

thf(icl_or,definition,
    ( icl_or
    := ( ^ [A: $i > $o, B: $i > $o] : (mor @ A @ B)) )).
         
%----ICL implication connective
thf(icl_impl,type,(
    icl_impl:  ($i > $o) > ($i > $o) > ($i > $o) )).

thf(icl_impl,definition,
    ( icl_impl
    = ( ^ [A: $i > $o, B: $i > $o] : (mbox @ rel @ (mimpl @ A @ B))) )).

%----ICL true connective
thf(icl_true,type,(
    icl_true:  ($i > $o) )).

thf(icl_true,definition,
    ( icl_true
    := mtrue )).

%----ICL false connective
thf(icl_false,type,(
    icl_false:  ($i > $o) )).

thf(icl_false,definition,
    ( icl_false
    := mfalse )).

%----ICL says connective
thf(icl_says,type,(
    icl_says:  ($i > $o) > ($i > $o) > ($i > $o) )).

thf(icl_says,definition,
    ( icl_says
    := ( ^ [A: $i > $o, S: $i > $o] : (mbox @ rel @ (mor @ A @ S))) )).

%%%%%%%%%%%%%%%%%%%%%%%%%%%%%%%%%%
% ICL notions of validity wrt. K %
%%%%%%%%%%%%%%%%%%%%%%%%%%%%%%%%%%

%----An ICL formula is K-valid if its translation into modal logic is valid
thf(iclval_decl,type,(
    iclval: ( $i > $o ) > $o )).

thf(icl_s4_valid,definition,
    ( iclval
    := ( ^ [X: $i > $o] : (mvalid @ X)) )).

\end{verbatim}
}

%\paragraph*{File ICL\_s4.ax}
The file ICL\_s4.ax provides the axioms \texttt{R} and \texttt{T} are added to 
to obtain a mapping into modal logic S4. 
{\scriptsize
%\verbatiminput{ICL\_s4.ax}
\begin{verbatim}
%------------------------------------------------------------------------------
% File     : ICL_s4.ax
% Domain   : ICL Logic and its translation into Modal Logic (which is
%            itself modeled in simple type theory; see [2])
% Axioms   : ICL logic based upon modal logic based upon simple type theory
% Version  : 
% English  :
% Refs     : [1] Deepak Garg, Martín Abadi: A Modal Deconstruction of Access 
%            Control Logics. FoSSaCS 2008: 216-230
%            [2] C. Benzmueller and L. Paulson. Exploring Properties 
%            of Normal Multimodal Logics in Simple Type Theory with LEO-II.
%            Festschrift in Honour of Peter B. Andrews. 
%            See: http://www.ags.uni-sb.de/~chris/papers/B9.pdf 
% Status   :
% Syntax   : 
% Comments : Formalization in THF by C. Benzmueller 
%------------------------------------------------------------------------------

%%%%%%%%%%%%%%%%%%%%%%%%%%%%%%%%%%
% ICL notions of validity wrt S4 %
%%%%%%%%%%%%%%%%%%%%%%%%%%%%%%%%%%

%----We add the reflexivity and the transitivity axiom to obtain S4.

thf(refl_axiom,axiom,
    (![A:($i>$o)]: (mvalid @ (mimpl @ (mbox @ rel @ A) @ A)) )).

thf(trans_axiom,axiom,
    (![B:($i>$o)]: (mvalid @ (mimpl @ (mbox @ rel @ B) @ 
                                      (mbox @ rel @ (mbox @ rel @ B )))) )).

\end{verbatim}
}

%\paragraph*{File ICL\_ex1\_s4.thf}
File  ICL\_ex1\_s4.thf contains the encoding of Example 1.
{\scriptsize
%\verbatiminput{ICL\_ex1\_s4.thf}
\begin{verbatim}
%------------------------------------------------------------------------------
% File     : ICL_ex1_s4.thf
% Domain   : ICL Logic and its translation into Modal Logic (which is
%            itself modeled in simple type theory; see [2])
% Axioms   : 
% Version  : 
% English  : ICL logic mapping to modal logic S4 implies 'Ex1'; see p.4 of [1]
% Refs     : [1] Deepak Garg, Martín Abadi: A Modal Deconstruction of Access 
%            Control Logics. FoSSaCS 2008: 216-230
%            [2] C. Benzmueller and L. Paulson. Exploring Properties 
%            of Normal Multimodal Logics in Simple Type Theory with LEO-II.
%            Festschrift in Honour of Peter B. Andrews. 
%            See: http://www.ags.uni-sb.de/~chris/papers/B9.pdf 
% Status   : Theorem (Henkin semantics)
% Syntax   : 
% Comments : Formalization in THF by C. Benzmueller 
%------------------------------------------------------------------------------

include('ICL_k.ax').
include('ICL_s4.ax').

%----The principals

thf(admin,type,(
    admin: $i > $o )).

thf(bob,type,(
    bob: $i > $o )).

%----The atomic propositions

thf(deletfile1,type,(
    deletefile1: $i > $o )).

%----The axioms of the example problem
%----(admin says deletefile1) => deletfile1
thf(ax1,axiom,
    (iclval @ 
       (icl_impl @ (icl_says @ (icl_princ @ admin) @ (icl_atom @ deletefile1)) 
                 @ (icl_atom @ deletefile1)) )).

%----(admin says ((bob says deletefile1) => deletfile1))
thf(ax2,axiom,
    (iclval @ 
       (icl_says @ (icl_princ @ admin) 
                 @ (icl_impl @ (icl_says @ (icl_princ @ bob) 
                                         @ (icl_atom @ deletefile1))
                             @ (icl_atom @ deletefile1))) )). 

%----(bob says deletefile1)
thf(ax3,axiom,
    (iclval @ (icl_says @ (icl_princ @ bob) @ (icl_atom @ deletefile1)) )).

%----We prove: It holds deletefile1
thf(ex1,conjecture,
    (iclval @ (icl_atom @ deletefile1) )).

\end{verbatim}
}

%\paragraph*{Files ICL\_unit\_s4.thf,  ICL\_cuc\_s4.thf,  ICL\_idem\_s4.thf}
Files ICL\_unit\_s4.thf,  ICL\_cuc\_s4.thf, and  ICL\_idem\_s4.thf contain the encodings of
the axioms unit, cuc and idem as proof problems.
{\scriptsize
\begin{verbatim}
%------------------------------------------------------------------------------
% File     : ICL_unit_s4.thf
% Domain   : ICL Logic and its translation into Modal Logic S4 (which is
%            itelf modeled in simple type theory; see [2])
% Axioms   : 
% Version  : 
% English  : ICL logic mapping to modal logic implies 'unit'; see p.3 of [1]
% Refs     : [1] Deepak Garg, Martín Abadi: A Modal Deconstruction of Access 
%            Control Logics. FoSSaCS 2008: 216-230
%            [2] C. Benzmueller and L. Paulson. Exploring Properties 
%            of Normal Multimodal Logics in Simple Type Theory with LEO-II.
%            Festschrift in Honour of Peter B. Andrews. 
%            See: http://www.ags.uni-sb.de/~chris/papers/B9.pdf 
% Status   : Theorem (Henkin semantics)
% Syntax   : 
% Comments : Formalization in THF by C. Benzmueller 
%------------------------------------------------------------------------------

include('ICL_k.ax').
include('ICL_s4.ax').

%----We introduce an arbitrary atom s
thf(s,type,(
    s : $i > $o )).

%----We introduce an arbitrary principal a
thf(a,type,(
    a : $i > $o )).

%----Can we prove 'unit'? 
thf(unit,conjecture,
    ( iclval @ (icl_impl @ (icl_atom @ s) 
                         @ (icl_says @ (icl_princ @ a) @ (icl_atom @ s))) )).


\end{verbatim}
\begin{verbatim}
%------------------------------------------------------------------------------
% File     : ICL_cuc_s4.thf
% Domain   : ICL Logic and its translation into Modal Logic S4 (which is
%            itelf modeled in simple type theory; see [2])
% Axioms   : 
% Version  : 
% English  : ICL logic mapping to modal logic implies 'cuc'; see p.3 of [1]
% Refs     : [1] Deepak Garg, Martín Abadi: A Modal Deconstruction of Access 
%            Control Logics. FoSSaCS 2008: 216-230
%            [2] C. Benzmueller and L. Paulson. Exploring Properties 
%            of Normal Multimodal Logics in Simple Type Theory with LEO-II.
%            Festschrift in Honour of Peter B. Andrews. 
%            See: http://www.ags.uni-sb.de/~chris/papers/B9.pdf 
% Status   : Theorem (Henkin semantics)
% Syntax   : 
% Comments : Formalization in THF by C. Benzmueller 
%------------------------------------------------------------------------------

include('ICL_k.ax').
include('ICL_s4.ax').

%----We introduce an arbitrary atom s and t 
thf(s,type,(
    s : $i > $o )).

thf(t,type,(
    t : $i > $o )).

%----We introduce an arbitrary principal a
thf(a,type,(
    a : $i > $o )).

%%----Can we prove 'cuc'? 
thf(cuc,conjecture,
    (iclval @ 
     (icl_impl 
      @ (icl_says @ (icl_princ @ a) @ (icl_impl @ (icl_atom @ s) 
                                                @ (icl_atom @ t)))	
      @ (icl_impl 
         @ (icl_says @ (icl_princ @ a) @ (icl_atom @ s))
         @ (icl_says @ (icl_princ @ a) @ (icl_atom @ t)))) )).
       
\end{verbatim}
\begin{verbatim}
%------------------------------------------------------------------------------
% File     : ICL_idem_s4.thf
% Domain   : ICL Logic and its translation into Modal Logic S4 (which is
%            itelf modeled in simple type theory; see [2])
% Axioms   : 
% Version  : 
% English  : ICL logic mapping to modal logic implies 'idem'; see p.3 of [1]
% Refs     : [1] Deepak Garg, Martín Abadi: A Modal Deconstruction of Access 
%            Control Logics. FoSSaCS 2008: 216-230
%            [2] C. Benzmueller and L. Paulson. Exploring Properties 
%            of Normal Multimodal Logics in Simple Type Theory with LEO-II.
%            Festschrift in Honour of Peter B. Andrews. 
%            See: http://www.ags.uni-sb.de/~chris/papers/B9.pdf 
% Status   : Theorem (Henkin semantics)
% Syntax   : 
% Comments : Formalization in THF by C. Benzmueller 
%------------------------------------------------------------------------------

include('ICL_k.ax').
include('ICL_s4.ax').

%----We introduce an arbitrary atom s and t 
thf(s,type,(
    s : $i > $o )).

%----We introduce an arbitrary principal a
thf(a,type,(
    a : $i > $o )).

%----Can we prove 'idem'?
thf(idem,conjecture,
    (iclval @ 
     (icl_impl 
      @ (icl_says @ (icl_princ @ a) @ (icl_says @ (icl_princ @ a) 
                                                @ (icl_atom @ s)))
      @ (icl_says @ (icl_princ @ a) @ (icl_atom @ s))))).
       
\end{verbatim}

\end{document}